\documentclass[aps,pre,twocolumn,floatfix,showpacs]{revtex4}
\usepackage{epsf,amsmath,amssymb}
\begin{document}
\title{Robustness of the avalanche dynamics in data packet
transport on scale-free networks}
\author{E.~J. Lee, K.-I. Goh, B. Kahng, and D. Kim}
\affiliation{School of Physics and Center for Theoretical 
Physics,\\
Seoul National University NS50, Seoul 151-747, Korea}
\date{\today}
\begin{abstract}
We study the avalanche dynamics in the data packet transport on
scale-free networks through a simple model. In the model, each 
vertex is assigned a capacity proportional to the load with the
proportionality constant $1+a$. 
When the system is perturbed by a single vertex 
removal, the load of each vertex is redistributed, followed by subsequent 
failures of overloaded vertices. 
The avalanche size depends on the parameter $a$ as well as 
which vertex triggers it. 
We find that there exists a critical value $a_c$ at which the 
avalanche size distribution follows a power law.
The critical exponent associated with it appears to be robust as long as the degree exponent is 
between 2 and 3, and is close in value to that of the distribution of the diameter 
changes by single vertex removal. 
\end{abstract}
\maketitle

Avalanche dynamics, triggered by small initial
perturbation, but spreading to other constituents successively, is
one of intriguing problems in
physics~\cite{watts,virus,newman,power,kinney,holme,motter,vespig,crucitti,bianconi,moreno}.
Such avalanche dynamics manifests itself in diverse forms such as
cultural fads~\cite{watts}, virus spreading~\cite{virus}, disease
contagion~\cite{newman}, blackout in power transmission
grids~\cite{power,kinney}, data packet congestion in the
Internet~\cite{holme,motter}, and so on. In particular, the avalanche
phenomena on complex networks are interesting, because they occur more 
frequently and their impact can be more severe than those occurring 
in the Euclidean space due to the close 
inter-connectivity among constituents in complex networks.

To understand the intrinsic nature of the avalanche dynamics 
on complex networks, the sandpile model proposed by Bak, Tang, and Wiesenfeld has 
been studied on scale-free (SF) networks recently~\cite{sandpile}. The SF 
network is the network whose degree distribution follows a power law, 
$p_d(k)\sim k^{-\gamma}$.  
Since the sandpile model is a self-organized critical model, 
the avalanche size distribution follows a power law, $p_a(s)\sim s^{-\tau}$, 
where $s$ is the avalanche size. 
In the sandpile model, the exponent $\tau$ depends on the degree exponent 
$\gamma$ of the embedded SF network as $\tau_{\rm BTW}=\gamma/(\gamma-1)$ 
for $2<\gamma<3$ when the toppling threshold of each vertex is equal to its 
degree. 
However, when the toppling threshold is fixed as a constant, 
independent of degree, the exponent $\tau_{\rm MF}=3/2$, 
being equal to the mean field value in the Euclidean space. 
Thus, it would be interesting to find an example of avalanche 
dynamics where the avalanche size distribution follows a power 
law with a nontrivial exponent, but different from the mean 
field value, and robust against variation of degree exponents. 
For this purpose, in this paper, we study the model proposed 
by Motter and Lai (ML)~\cite{motter}, designed to exploit 
the avalanche dynamics in the process of data packet transport 
on complex networks.

In the ML model, each vertex is assigned a finite capacity, given as 
\begin{equation}
c_j=(1+a)\ell_j^{(0)},
\label{cap}
\end{equation}
where $a$ is a control parameter and $\ell_j^{(0)}$ is the
load of vertex $j$. The load of a given vertex is defined as the 
sum of the effective number of data packets passing through 
that vertex when every pair of vertices send and receive a unit 
data packet. The data packets are allowed to travel along the shortest 
pathways between a given pair of vertices and are divided evenly 
at each branching point~\cite{load, KI}. For SF networks, 
the load of each vertex is 
heterogeneous, and its distribution also follows a power law, 
$p_{\ell}(\ell)\sim \ell^{-\delta}$. 
The superscript $(0)$ in Eq.~(1) indicates the load without any 
removal of vertices. 
The excess term $a\ell^{(0)}_j$ in Eq.~(1) provides the ability
to tolerate the additional burden and may describe the excess buffer
at the routers in the Internet, for example.
The basic assumption of the ML model is that the size of
such an excess buffer is proportional to the activity at the vertex,
the load $\ell^{(0)}_j$. The control parameter $a$ sets the
global level of tolerance of the system.

Next, we remove a vertex $i$ intentionally, 
which we call triggering vertex.
Then each pair of remaining vertices whose shortest pathway had
passed through that triggering vertex should find detours, resulting
in rearrangement of the shortest pathways over the network, and
the load at a remaining vertex $j$ takes a new value, which is
denoted as $\ell_j^{(i)}$. If the load $\ell_j^{(i)}$ exceeds its
capacity $c_j$ given by Eq.~(\ref{cap}), then the vertex $j$ would
fail irreversibly. Other overloaded vertices also fail at the
same time. These are the failures by the first shock,
marked green with the symbol ($\diamond$) in Fig.~1.  
After then, the shortest pathway configurations would rearrange again, 
and the overloaded vertices fail successively until no overloaded 
vertices remain. The avalanche size $s_i$ is defined as the total 
number of failed vertices throughout the whole process of the avalanche
triggered by the vertex $i$. 
Note that in this model, failures do not necessarily proceed 
contiguously, that is, through the neighbors of vertices previously failed, 
but spread over the entire system through nonlocal dynamics as 
shown in Fig.~1. For such nonlocal dynamics, the
branching process formalism cannot be used to obtain the avalanche 
size distribution of the ML model.
\begin{figure}[t]
\centerline{\epsfxsize=5.5cm \epsfbox{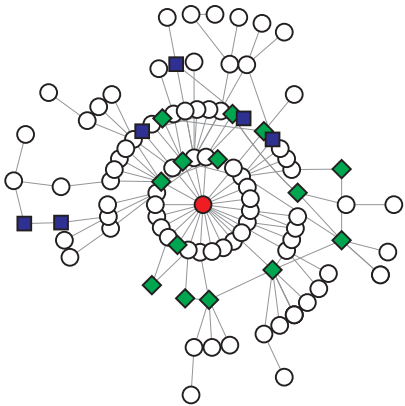}}
\caption{(Color online) Plot of the avalanche dynamics pattern at 
$a_c=0.15$ for a given small-size network. 
Cascading failures starting from the central vertex spread in a nonlocal way following the 
steps, $\bigcirc$ (red), $\Diamond$ (green), and $\Box$ (blue).} 
\end{figure}

In the original work, ML measured the ratio $G_i=N_i'/N$, where $N$ 
and $N_i'$ are the numbers of vertices before and after cascading 
failures, respectively, when the triggering vertex is $i$. 
Note that the avalanche size corresponds to $s_i=N-N_i'$. 
ML found that $G_i$ depends on the degree $k_i$ of the triggering 
vertex $i$ as well as the control parameter $a$. 
When $a$ is large (small), the capacity of each vertex is large 
(small), so that the number of failed vertices is small (large) and 
$G_i$ is close to one (zero). Moreover, when the degree of 
the triggering vertex is large (small), $G_i$ is close to zero (one), 
and the system is vulnerable (robust). Such numerical results 
suggest that there may occur a phase transition in the avalanche size. 
In this paper, we find numerically that indeed there exists a critical 
value $a_c$ at which the avalanche size distribution follows a power 
law, $p_a(s)\sim s^{-\tau}$. 
We also study various features of the avalanche dynamics at the 
critical point. 
\begin{figure}[t]
\centerline{\epsfxsize=8.cm \epsfbox{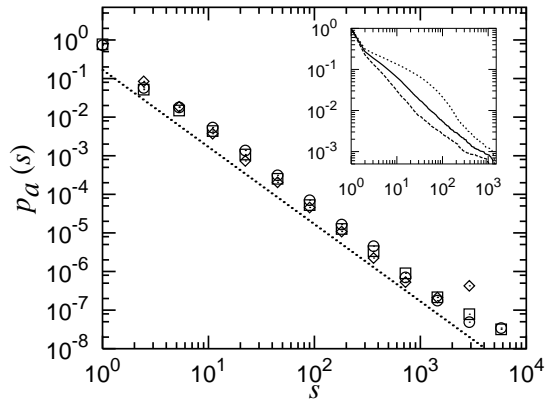}}
\caption{Plot of the avalanche size distribution for the BA model
at $a_c=0.15$ with different $\gamma=3.0$ ($\bigcirc$), 
2.6 ($\Box$), and 2.2 ($\diamond$). The mean degree is 4, and the system size is $N=10^4$. 
The dotted line has a slope $-2.1$, drawn for reference. 
The avalanche size distributions are obtained by deleting each vertex
$i$ in turn and measuring the respective avalanche size $s_i$,
then tabulating the histogram of $s_i$, normalized by the number of
triggering vertices $N$.
Inset: the avalanche size distribution(cumulative) under the same condition for $\gamma=3.0$, 
but with $a=0.11$ (top), 0.15 (middle), 
and 0.2 (bottom).} 
\label{ps_motter}
\end{figure}
\begin{figure}
\centerline{\epsfxsize=7.8cm \epsfbox{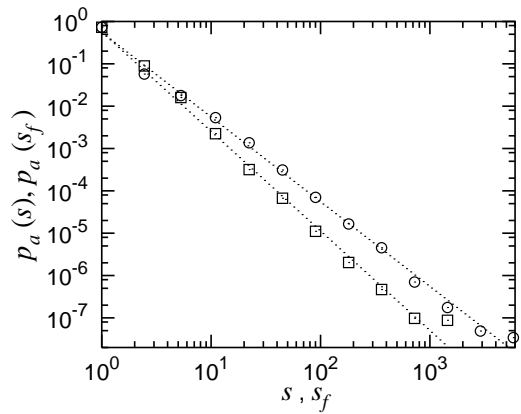}}
\caption{Plot of the avalanche size distribution for the BA 
model with $\gamma=3$ by the first shock ($\Box$), compared 
with the avalanche size distribution including the 
entire process ($\bigcirc$). The slopes of dotted and dashed lines 
are $-2.3$ and $-2.1$, respectively, drawn for reference.} 
\label{fig:first_shock}
\end{figure}
\begin{figure}[t]
\centerline{\epsfxsize=8.cm \epsfbox{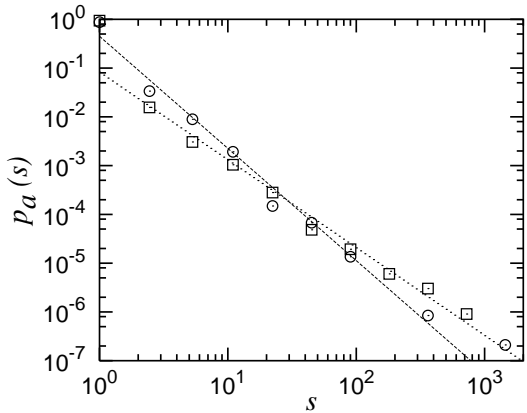}}
\caption{Plot of the avalanche size distribution for the yeast 
protein interaction network ($\bigcirc$) and the Internet ($\Box$) at $a_c=0.15$. 
The slopes of dotted and dashed lines are $-1.8$ and $-2.3$, respectively, 
drawn for reference.}  
\label{yeast_int_adist}
\end{figure}

Let us first investigate the distribution of $\{s_i\}$, the avalanche size distribution $p_a(s)$.
For large (small) $a$, the number of overloaded vertices is small 
(large), so that the avalanche size is finite (diverges) and the system may be
considered  as in a subcritical (supercritical) phase. 
We find that there exists a characteristic value $a_c$ between the two regimes, 
where the avalanche size distribution follows a power law,
$p_a(s)\sim s^{-\tau}$ as shown in Fig.~\ref{ps_motter}.
Numerical simulations are performed for the Barab\'asi-Albert (BA)
model~\cite{ba} with different degree exponent values. 
We find that $a_c \approx 0.15$, and
$\tau\approx 2.1(1)$, both of which are likely to be robust for different 
degree exponents $\gamma$ as long as $2 < \gamma < 3$. 
While it is not manifest why such a robust behavior
occurs in the avalanche size distribution, it is noteworthy to remind
that other problems related to the shortest pathways such as
the load distribution and the diameter change distribution are also
likely to be robust. Thus the robustness of the avalanche size distribution
may be caused by the notion of the shortest pathway.
For $\gamma > 3$, however, $p_a(s)$ decays with exponent 
larger than $\tau \approx 2.1$ or exponentially depending on $\gamma$. 
The avalanche size distribution by the first shock 
behaves differently as $p_a(s_f)\sim s_f^{-2.3}$, which is shown in 
Fig.~3. We also check the avalanche size distribution for real world
networks. 
For the yeast protein interaction network and the Internet,
we obtain $\tau \approx 2.3(1)$ and $\tau \approx 1.8(1)$, 
respectively, as shown in Fig.~\ref{yeast_int_adist}. Note that 
the degree exponent of the yeast protein interaction network 
is $\gamma\approx3.4$~\cite{hybrid}, slightly larger than $3$,
thus the exponent $\tau\approx 2.3$ is somewhat larger than $2.1(1)$ 
obtained in the BA model for $2 < \gamma <3$. 

The deviation of the exponent $\tau$ for the Internet
is rooted from its different network structure from those of the protein
interaction networks or the BA-type model networks: it was
found through recent several studies that the Internet structure is 
effectively tree-like, while the protein interaction network
and BA-type model networks 
contain diverse connections \cite{KI,dhkim}.
Accordingly, when a vertex on a branch of tree structure is
removed, the giant cluster is divided into two or more components,
and the giant cluster size shrinks apparently. Such a case occurs
more often in the Internet than in other-type networks,
because the Internet is tree-like. Due to this fact, the avalanche
size statistics of the Internet is different from that of other networks.
On the other hand, 
while the rule of the ML model may not be relevant to the dynamics
in the protein interaction network or cellular networks, 
cascading failure occurring in cellular
networks is an important concept. For example, the protein interaction
network provides
the basic operational protocol in various signal transduction and functional
pathways. In such a system, when a certain element (a protein or 
a substrate) fails or is removed (perturbed), others should take over
its burden to survive the lack thereof, although the mechanism by which
the cascade spreads could be different from that of the ML model.
Studies in this direction have recently been carried out for the
metabolic networks \cite{lemke,cmghim}.

\begin{figure}
\centerline{\epsfxsize=8.0cm \epsfbox{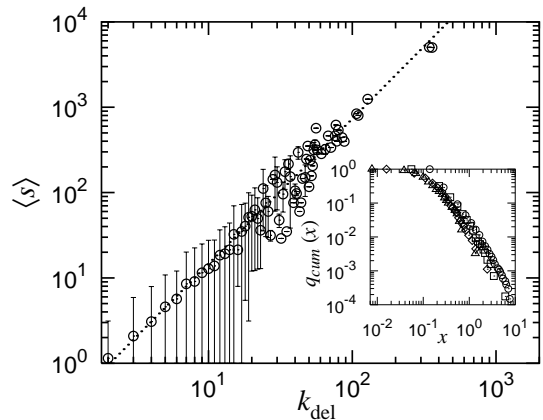}}
\caption{Plot of the mean avalanche size $\langle s \rangle$ 
versus the degree of the triggering vertex $k_{\rm del}$ for the BA model ($\gamma=3$)
at $a_c$. Data points ($\bigcirc$) are averaged over different 
avalanche sizes triggered by the vertices with a given $k_{\rm del}$. 
The standard deviation of each data point is represented by a bar. 
The slope of the dotted line is the theoretical value $1.8$, drawn 
for reference. Inset: Cumulative plot of the avalanche size distribution 
with the rescaled quantity $x=s/k_{\rm del}^{(\gamma-1)/(\tau-1)}$ 
for $k_{\rm del}=3$ ($\circ$), 5 ($\Box$), 10 ($\diamond$), 
and 15 ($\triangle$). }  
\label{s_k}
\end{figure}

Next, we examine the relationship of the mean avalanche size, 
denoted by $\langle s \rangle(k_{\rm del})$, over different triggering vertices 
but with a given degree $k_{\rm del}$ at $a_c$ in Fig.~\ref{s_k}. 
We find that the quantity $\langle s \rangle(k_{\rm del})$ increases with 
increasing $k_{\rm del}$. However, there occur large fluctuations 
in $\langle s \rangle(k_{\rm del})$, in particular, for small $k_{\rm del}$. 
Note that if the orderings in the magnitude of 
$\langle s \rangle(k_{\rm del})$ and $k_{\rm del}$ are preserved, 
$p_a(s){\rm d}s=p_d(k){\rm d}k$ and hence one has the relation 
$\langle s \rangle(k_{\rm del}) \sim k_{\rm del}^{(\gamma-1)/(\tau-1)}$. 
Indeed, Fig.~\ref{s_k} exhibits such a behavior.  
To examine the fluctuations of $\langle s \rangle(k_{\rm del})$ for given 
$k_{\rm del}$, we consider the distribution function $q(x)$ of the 
avalanche sizes for given $k_{\rm del}$ with a rescaled 
quantity, $x=s/k_{\rm del}^{(\gamma-1)/(\tau-1)}$.
Shown in the inset of Fig.~\ref{s_k} are the data of the cumulative distribution of $q_{\rm cum}(x)$ for 
different $k_{\rm del}$, which collapse onto a single curve exhibiting a fat-tail 
behavior as $q(x)\sim x^{-3.2}$ for large $x$. 

\begin{figure}[t]
\centerline{\epsfxsize=7.5cm \epsfbox{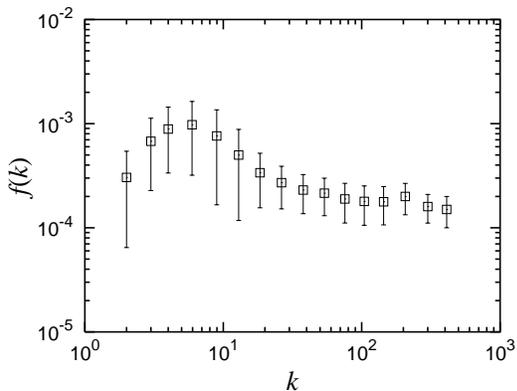}}
\caption{Plot of the failure fraction $f$ versus degree $k$ at 
$a_c$ for the BA model ($\gamma=3$) with $N=10^4$. Data points are
logarithmically binned. 
Error bars represent the standard deviations for each bin.}
\label{fig6}
\end{figure}

Next, to study how much a given vertex with degree $k$ is vulnerable 
or robust under a random vertex failure, we count the number 
of failures $n_j$ of a vertex $j$ out of $N$ cascading events 
when each of $N$ vertices acts as the triggering vertex.
At this point, it is convenient to consider the random variables 
$x_{j}^{\phantom{i}i}$ which
take the value 1 if the vertex $j$ topples due to the triggering vertex $i$ 
and 0 otherwise. In terms of $x_{j}^{\phantom{i}i}$, 
$\sum_{j}^{}{ x_{j}^{\phantom{i}i} }=s_i$ and $\sum_{i}{ x_{j}^{\phantom{i}i}}
=n_j$.
Let $f(k)$ be the average of $n_j/N$ over the vertices with degree $k$. 
Fig.~\ref{fig6} shows the function $f(k)$ versus $k$. 
It increases with increasing $k$ for small $k$ 
and exhibits a peak in the intermediate range of $k$. 
For large $k$, $f(k)$ is almost independent of $k$.
This result implies that the vertices with degree in the intermediate 
range are more vulnerable. 
\begin{figure}
\centerline{\epsfxsize=7.0cm \epsfbox{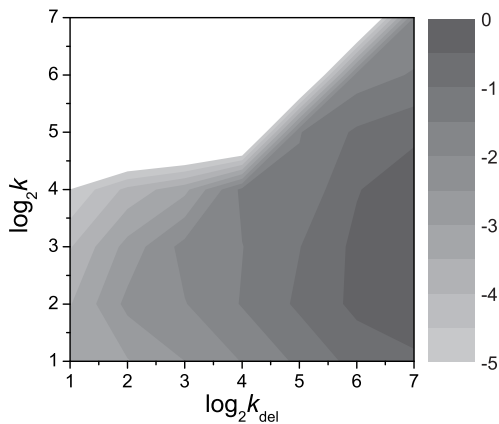}}
\caption{Plot of the logarithm (with base 10) 
of the failure correlation function 
$c(k,k_{\rm del})$ as functions of the degrees of the failed vertex 
$k$ and of the triggering vertex $k_{\rm del}$. 
Data are logarithmically binned to reduce fluctuations.
Simulation is performed for the BA model with $\gamma=3$,
$N=3000$ and averaged over 10 configurations.}  
\label{fig7}
\end{figure}
Meanwhile, 
we note that the asymptotic value of $f(k)$ for large $k$ is 
$\sim {\cal O}(1/N)$. This is because a vertex with large $k$
hardly fails through the cascading failure process triggered by other
vertices, but fails trivially when itself acts as the triggering vertex. 
The peak at the intermediate value of $k$ in $f(k)$ can be understood
in the following heuristic way. The load itself quantifies the level
of traffic coming through the vertex, so we can expect that the higher
the load is, the more excess traffic it would get by the breakdown
of other vertices. On the other hand, since the excess capacity 
$a\ell_j^{(0)}$ is assigned in a multiplicative way, the higher  the load
is, the larger room to accommodate the excess is, reducing the occasion
to be overloaded. These two factors compete each other,
generating a peak in the intermediate range of $k$ in $f(k)$. That implies
the vertices with intermediate degrees are more vulnerable. Such
a behavior can also be seen in the information cascade model of Watts \cite{watts}.

This result is also reminiscent of the avalanche 
dynamics of the sandpile model.
The hubs, vertices with large degrees, 
play a role of the reservoir against failures~\cite{sandpile}.
We also consider the failure correlation function $c(k,k_{\rm del})$, 
defined as the average of $x_j^{\phantom{i}i}$ with the constraints 
$k_j=k$ and $k_i=k_{\rm del}$, 
$k_i$ denoting the degree of a vertex $i$.
The darkest region in the bottom-right corner of Fig.~7 indicates
that vertices with small degrees easily fail by the trigger of vertices
with large degrees, whereas the reverse rarely happen, particularly
for large $k_{\rm del}$, as manifested by the white region in the upper-left
part of Fig.~7.

It is interesting to notice that the avalanche size distribution 
behaves similarly to the diameter change distribution~\cite{diameter_change}. 
Diameter is the average number of hops between every pair of vertices. 
Let $d^{(0)}$ be the diameter of a given network, where 
the superscript ${(0)}$ means unperturbed network. 
When the network is perturbed by the removal of a vertex $i$, 
the diameter changes accordingly, and the diameter 
of the remaining network is denoted as $d^{(i)}$.
Then the dimensionless quantity $\Delta_i=(d^{(i)}-d^{(0)})/d^{(0)}$ 
is measured for all $i$, and then its distribution function, composed 
of $\{\Delta_i\}$, behaves as $p_{\rm DC}(\Delta)\sim \Delta^{-\zeta}$ 
for large $\Delta$. The exponent $\zeta$ was measured to be 
$\zeta\approx 2.2(1)$ 
for most artificial SF networks including the BA model, insensitive to 
the degree exponent $\gamma$ as long as $2 < \gamma < 3$, 
and $\zeta\approx 2.3(1)$ for the yeast protein interaction network, 
but $\zeta \approx 1.7(1)$ for the Internet. All the above values 
of the exponent $\zeta$ are close to corresponding values of 
$\tau$ for the avalanche size distribution of the ML model.
In addition, the exponents $\tau$ and $\zeta$ are also close 
in values to the load distribution exponent $\delta$ except for a few 
examples such as the Internet. Thus, it would be interesting 
to investigate the origin of such coincidences on a fundamental 
level. 

Finally, it is noteworthy that recently Zhao {\it et al.}~\cite{zhao} 
also studied the phase transition of the cascading failure for the 
ML model. They estimated the critical point to be $a_c\approx 
0.1$ by comparing the load distribution before and after the 
deletion of the hub. Their estimation is not inconsistent 
with our numerical estimation. However, the avalanche 
size distribution studied in this work provides a better criterion
for the phase transition point.

In conclusion, we have studied the avalanche dynamics in the model 
proposed by Motter and Lai, describing the data packet transport 
on SF networks. Depending on the model parameter $a$, which controls 
the magnitude of the capacity of each vertex, the pattern of avalanche 
dynamics can change. For small $a$, cascading failure spreads over 
the entire system, corresponding to supercritical behavior 
in avalanche dynamics. While, for large $a$, cascading failure is 
confined in a small region, and avalanche size follows a subcritical 
behavior. At the critical point $a_c$, the avalanche size distribution 
follows a power law with exponent $\tau$. 
The exponent $\tau$ seems to be robust for different degree 
exponent $\gamma$ as long as $2 < \gamma < 3$, and is 
likely to be close to the exponent of the diameter change 
distribution.

This work is supported by the KOSEF grants No.
R14-2002-059-01000-0 in the ABRL program.


\end{document}